\documentstyle[epsfig, twocolumn]{esapub}

\begin{document}

%% Do not remove the following six lines:
\setlength{\parindent}{0pt}
\setlength{\parskip}{ 10pt plus 1pt minus 1pt}
\setlength{\hoffset}{-1.5truecm}
\setlength{\textwidth}{ 17.1truecm }
\setlength{\columnsep}{1truecm }
\setlength{\columnseprule}{0pt}
\setlength{\headheight}{12pt}
\setlength{\headsep}{20pt}
\pagestyle{esapubheadings}

%% Title - should be in capitals:
\title{\bf MULTIFREQUENCY AND IUE CAMPAIGNS ON BLAZARS}
%\title{\bf A \LaTeX{}-BASED FORMAT FOR PROCEEDINGS\thanks{these instructions
%        should be used to prepare manuscripts for the published proceedings}}

%% If the author list spans more than one line then the {\bf (bold
%% font)} command must be inserted for each line
\author{{\bf L. Maraschi$$} \vspace{2mm}\\
$$ Brera Astr. Obs., Via Brera 28, I-20121 Milan, Italy}

\maketitle

\begin{abstract}
The IUE contribution to the understanding of the blazar phenomenon has been
of fundamental importance. Here I review the progress obtained with
the latest multifrequency campaigns performed with IUE on two 
prototype objects, the BL Lac PKS 2155 - 304 and the  highly polarized,
superluminal quasar 3C 279.
\vspace {5pt} \\

%% Do not remove the previous commands. Your abstract should 
%% end with \vspace {5pt} \\  

%% Please insert your keywords here.
  Key~words: ultraviolet; blazar continuum; blazar emission
mechanisms; jets.

\end{abstract}

\section{INTRODUCTION}

The unique capabilities of IUE in spectrophotometric accuracy and UV sensitivity   
made it an essential tool in the study of variable X-ray sources. In particular 
for blazars it allowed to determine the shape of the non-thermal continuum 
in a region free of the possible contribution from a host galaxy and,
 together  with coordinated optical and X-ray 
observations, made possible by the flexible scheduling,
over a very wide spectral range.
It is worthwhile to recall here that the first evidence of a sytematic
difference in
spectral shape between X-ray selected BL Lacs and other blazars and "normal" 
quasars was obtained with IUE (Ghisellini et al. 1986). Systematic analyses
of IUE data concerning blazar variability are presented by Treves and 
Girardi (1990),  Edelson (1992), Pian and Treves (1993).
Early results from blazar observations with IUE
 are reviewed in Bregman, Maraschi and Urry (1987). 

It became clear after the results of the first pioneering studies 
that at least some
of the sources were varying extremely rapidly also in the UV and that
quasi - simultaneous snapshots of the UV to X-ray 
energy distribution could describe 
some average "state" but were insufficient to
probe the correlation between the two wavelength ranges. Simultaneous
 light curves in X-rays and UV and possibly other wavelengths 
were  and are
needed to address physical models of variability. In fact, 
despite the long lifetime of IUE well sampled multiwavelength data were obtained
in a limited number of cases (see for reviews Wagner \& Witzel 1995;
 Ulrich, Maraschi, \& Urry 1997, UMU97 hereafter).
 Here I will discuss two sources for which many data
have been obtained which are helping us to make progress in the understanding
of the blazar phenomenon. The first is the BL Lac object PKS 2155-304, one of the
brightest blazars in the UV and soft X-ray sky (Section 3). The second 
is the superluminal quasar  3C 279, the first and one of the brightest blazars
detected in $\gamma$-rays (Section 4). Before  discussing two apparently
different cases I will briefly present a scheme for a unitary phenomenological
and theoretical understanding of the non thermal emission of blazars.

\section{A UNITARY VIEW OF THE BLAZAR PHENOMENON}

The extreme properties of some radio loud AGN regarding rapid high amplitude
variability implying "strong" violations of the brightness temperature limit
($T_b < 10^{12} K$ to avoid a Compton catastrophe) and/or of the efficiency
limit ($\Delta L / \Delta t < 10^{42}$ erg s$^{-2}$) led to the hypothesis that 
 bulk relativistic effects were involved (Blandford \& Rees 1978).
Soon the idea was supported by the observation of superluminal motion
and presently it is almost universally accepted that the continuum
emission from blazars derives from a relativistic jet 
seen at small angle to its axis (e.g., Urry \&  Padovani 1995, UP95
hereafter). In practice
all strong radio sources with flat radio spectra and core dominant morphologies
are thought to be blazars.

The signature of "thermal" emission in the form of broad emission lines 
and in very few cases of a UV bump are present in some blazars (flat spectrum
radio quasars, FSRQ), 
while lineless objects, called BL Lacs, are usually considered as a
separate class despite the similar (but not identical) continuum 
properties. A further subdivision within BL Lacs arose from the selection
criterion, either in the radio or in the X-ray band (RBL --XBL). 
RBLs tend to have a much lower X-ray to radio flux ratio than XBL
and the question is how representative are these samples with respect to the
full population (e.g., UP95).
A comparison of the radio to X-ray continua of XBL, RBL and flat spectrum
radio quasars (FSRQ) (Sambruna et al. 1996) suggests that in all cases 
the radio to UV emission is synchrotron radiation.
For XBL the X-ray continuum falls on the extrapolation of the optical UV 
continuum and is on average steeper, suggesting a synchrotron origin, while
in  FSRQ the X-ray spectrum is flatter and in excess of an 
extrapolation from lower frequencies, suggesting the onset of a different 
spectral component, most probably  inverse Compton. From the point of
view of the continuum shape RBL appear intermediate between XBL and FSRQ.
 
The recent discovery by EGRET on board the Compton Gamma-ray Observatory
of $\gamma$-ray emission from blazars revealed that a substantial fraction
 and in many cases most of the power is 
emitted in this very high energy band, which therefore appears to play 
a fundamental role.
It is  important to ask whether the $\gamma$-ray emission is a general 
property of the whole blazar class or whether it differs in
different objects and subclasses. We have recently addressed this problem  
(Fossati et al. 1997) collecting and studying
multifrequency data (including $\gamma$-rays) for three complete samples 
of blazars: the 2 Jy sample of FSRQs (UP95, and
references therein), the 1 Jy sample
of BL Lac objects (Stickel, Meisenheimer, \& K\"uhr 1994) and the sample
of BL Lacs selected in the X-ray band from the Einstein Slew Survey 
(Perlman et al. 1996).
Although the percentages of objects detected in $\gamma$-rays are not large
and are different in different samples (40\%, 26\% and 17\%, respectively)
a consistent picture seems to emerge from the average spectral
energy distributions (SED) as shown in Figure 1.
In this figure all blazars in the three complete samples were merged 
and binned according to radio-luminosity irrespective of the original 
classification. The dashed lines derive from an analytic parametric
approximation based on simple assumptions. 
The main results are the following:
\begin {itemize} 

\item
for all luminosity classes the SEDs can be described by two 
quasiparabolic components and exhibit two broad peaks;

\item
both peaks  fall at lower frequencies for the higher luminosity
objects;

\item 
the peak frequency of the higher energy component correlates with the
peak frequency of the first one in such a way that their ratio could be
constant, as was assumed  for the dashed curves.

\end{itemize} 

The dashed curves further assume that the peak luminosity of the second
component is proportional to the 5 GHz radio luminosity. 

\begin{figure}[h]
  \begin{center}
    \leavevmode
  \centerline{\epsfig{file=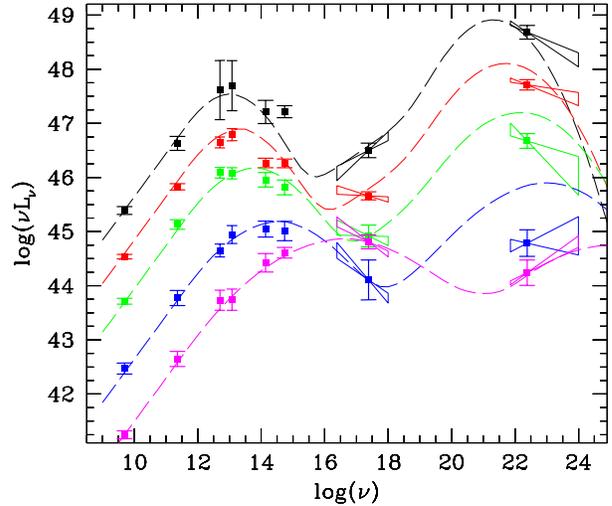,height=8cm}}
  \vspace{-1.8cm}
  \end{center}
  \caption{\em Average SEDs from radio to $\gamma$-rays for the "total blazar sample"
(see text) binned according to radio luminosity, irrespective of the original
classification. Empty asymmetric triangles represent uncertainties in spectral shapes
as measured in the X-ray and $\gamma$-ray bands.
The overlayed dashed curves are analytic approximations obtained
assuming that the ratio of the two peak frequencies is the same for all objects and
that the luminosity at the second peak is proportional to the radio luminosity
(from Fossati et al. 1997). }
%\label{fig:sample1}
\end{figure}

Although one should be aware  that selection biases may affect the results 
(as discussed in Fossati et al. 1997), Figure 1 suggests that the SEDs
of all blazars are globally similar and lie along a continuous
spectral sequence.  
For the most luminous objects the first peak falls at frequencies lower
than the optical band, while for the least luminous ones the reverse is true.
Thus highly luminous objects have  steep ("red") optical-UV continua,
while low luminosity objects with peak frequency beyond the UV range
have flat ("blue") optical-UV continua. For brevity I will refer
to objects on the high and low luminosity end of the sequence 
 as "red" or "blue" blazars.

It is generally thought that the first spectral component is due to 
synchrotron radiation. The spectra from the radio to the submillimeter range
most likely involve superposed contributions from different regions of the jet 
with different self-absorption turnovers. From infrared (IR)
frequencies upwards
the synchrotron emission should be thin and could be produced in a single
zone of the jet, allowing to adopt a homogeneous model.
The high energy electrons emitting the thin synchrotron radiation
could also produce the second spectral component (peaking in $\gamma$-rays)
by upscattering soft photons via the inverse Compton process. The seed photons
for upscattering could be the synchrotron photons themselves 
(synchrotron self Compton model, SSC) or photons outside the jet
(external Compton model, EC) possibly produced in an accretion disk
or torus and scattered or reprocessed by the surrounding gas
(e.g., Sikora 1994; UMU97, and references therein).

If the same region is responsible for the two spectral components,
 irrespective of the nature of the seed photons, {\it  emission
at the two peaks must derive from the same high energy electrons}.
Therefore a change in the density and/or spectrum of those electrons 
{\it is expected to cause correlated variability at frequencies 
close to the two peaks.} In the SSC model the inverse Compton
intensity is expected to vary more than the synchrotron one (due to the
same electrons), approximately as the square of it in the simplest case.
In the EC model, if the seed photons outside the jet are not
coupled to the jet emission, one expects a linear relation
between the synchrotron and inverse Compton variation.

Measuring the two peaks simultaneously is thus an essential step for
determining the physical parameters of the emission region and studying the
variability of the spectra around the peaks  yields unique insight
into the mechanisms of particle acceleration and energy loss in the
jet. The variability correlation should enable to disentangle the
contribution of different sources of seed photons (SSC vs. EC).

The "spectral sequence" discussed above could be attributed to a systematic
dependence of the critical electron energy (the break energy)
and/or the magnetic field or other properties of the jet on luminosity.
Assuming that the beaming factors are not significantly different
along the sequence, the trend in apparent luminosity is also
a trend in intrinsic luminosity. In the SSC model the break energy of the 
electrons (those radiating at the peaks) is fixed by the ratio of the two
peak frequencies and should therefore be approximately constant.
"Red" blazars should then have lower magnetic fields than "blue" blazars.
Systematic model fitting of all the $\gamma$-ray detected blazars with
sufficient multifrequency data suggests that as the magnetic field 
decreases the external photon density becomes important
so that a smooth transition between the SSC and EC scenario 
takes place. 

For "blue" blazars the UV range falls at frequencies lower 
than the synchrotron peak (to the left of it).
Adopting a homogeneous model the UV emission is therefore
attributed to synchrotron radiation by electrons 
of lower energy than those producing the X-rays (by the same mechanism).

For "red" blazars the UV falls at frequencies higher than 
(to the right of) the synchrotron peak. Also in this case the UV 
emission should be due to the synchrotron mechanism, but it should derive
from the highest energy electrons present. Rapid variability is therefore
expected.  
On the contrary, for the same type of objects, X-rays should be attributed
to inverse Compton of electrons of relatively lower energy, less
rapidly variable. 

Most of the known blazar phenomenology can be at least qualitatively
understood in this scheme (e.g., UMU97).
The multifrequency campaigns discussed below concern one "blue" (PKS 2155-304)
and one "red" (3C 279) blazars.

\section {PKS 2155-304}

PKS 2155-304 is one of the brightest sources in the extragalactic sky at 
UV and soft X-ray wavelengths. As such it was repeatedly observed in many
wavebands (Maraschi et al. 1986; Urry et al. 1988;
Treves et al. 1989; Edelson et al. 1991).
Two major intensive multiwavelength campaigns were organized on this source.
The first was based on 5 days of quasicontinuous coverage with IUE and 
3.5 days  with ROSAT (Urry et al. 1993, Brinkmann et al. 1994, 
Edelson et al. 1995).
The UV and X-ray light curves showed rapid variations of moderate amplitude 
($\simeq 20\%$) strongly correlated in the overlapping time intervals with the
soft X-rays leading the UV variations by not more than 2-3 hours.
Therefore variability was essentially "achromatic", that is independent of 
wavelength, ruling out an accretion disk as the origin of the UV continuum.
Over the whole monitoring period of about 1 month the UV intensity increased
by a factor 2, the same as observed in the optical and IR. 

A second, longer campaign, with $\simeq$ 10 days of IUE, $\simeq$ 9 days of EUVE,
2 days of ASCA and three short ROSAT observations took place in May 1994 
(Pesce et al. 1997, Pian et al. 1997, Urry et al. 1997).
The light curves from this campaign are shown in Figure 2. The source cooperated
nicely producing a well defined flare (factor 2.5) during the two days of ASCA 
observations
which can be most plausibly related with the smaller amplitude intensity peaks 
(35~\%)
seen at later times in the EUVE and IUE light curves.  The X-ray flare 
appears to lead
the EUVE and UV events by 1 and 2 days respectively, an order of magnitude 
longer than
the lag detected in the previous campaign. Within the ASCA data the 0.5 -- 1 keV 
photons lagged the 2.2 -- 10 keV photons by 1.5 hours. 

\begin{figure}[h]
  \begin{center}
    \leavevmode
  \centerline{\epsfig{file=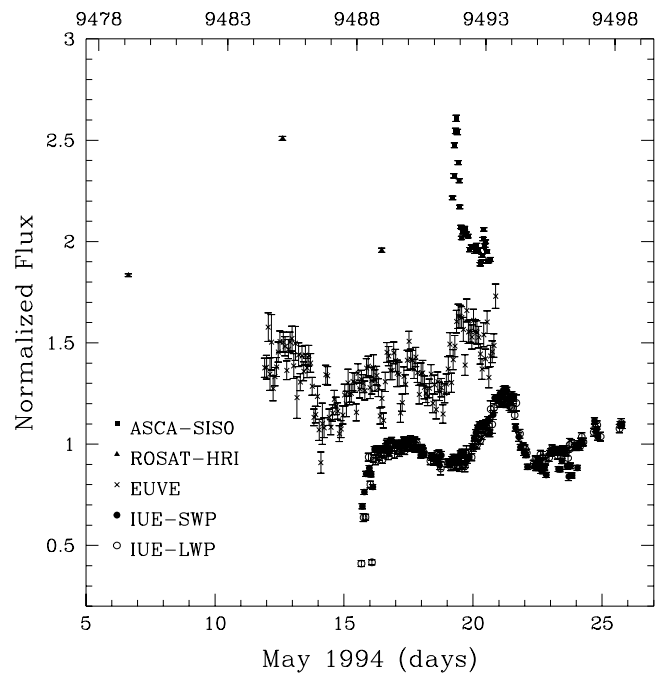,height=9cm}}
%  \vspace{1.cm}
  \end{center}
  \caption{\em Normalized X-ray, EUV and UV 
 light curves of PKS 2155-304 from the
second intensive multiwavelength monitoring campaign, in May 1994. The ASCA data show
a strong flare, echoed one day later in the EUVE and two days later in the
IUE light curves. The amplitude of the flares decreases and the duration increases
with increasing wavelength (from Urry et al. 1997).}

%\label{fig:sample1}
\end{figure}

Note how the UV light curve is well defined and how the differences between 
the LWP and SWP normalized fluxes are small, except for the initial 
extremely fast
and deep minimum. This event is discussed in detail in Pian et al. (1997) 
and is not understood. 

Despite the differences these two campaigns give the first evidence that the 
variations from 5 eV to 10 keV are correlated on short time scales and
that high frequencies lead the lower ones. 
The flare event observed during the second campaign has characteristics as 
expected from the propagation of a shock wave along an inhomogeneous 
relativistic jet. The cause of the flare in this model is a propagating
disturbance affecting different regions at different times.

However homogeneous models are perhaps more appealing as they can be 
strongly constrained by the data. The apparent progression of the X-ray 
flare to longer wavelengths rules out a stochastic acceleration 
process in the emitting region. It can possibly be explained by an
 "instantaneous" injection of high energy particles near X-ray emitting
energies and subsequent energy loss by synchrotron radiation in a homogeneous
region. Interpreting  the observed lags as radiative cooling times at the
relevant energies yields a univocal value of the magnetic field
B = 0.1 G. This rather low value of B requires a high value of the Doppler 
beaming factor ($\delta \simeq 30$) in order to avoid overproduction of 
$\gamma$-rays by the inverse Compton process. We recall however
that the $\gamma$-ray flux was not measured simultaneously.

Although the case of PKS 2155-304 is the best studied from X-ray to UV
frequencies, a closely similar behaviour was observed also in Mkn 421 
(Macomb et al. 1995; Buckley et al. 1996; Takahashi et al. 1996). Much more
extreme events, though probably
related to similar physics were recently discovered in Mkn 501 
(Pian et al. 1998a).

\section{3C 279}

 3C~279 ($z = 0.54$)  was the first blazar discovered by EGRET on board the
Compton Gamma-Ray Observatory to emit strong and variable $\gamma$-rays (June 1991).
This is  a "red" blazar where high energy electrons in a relativistic jet emit
 synchrotron radiation peaking between sub-mm and IR frequencies.
  Inverse Compton scattering of the same electrons  off synchrotron photons in the
jet (SSC) or on "ambient" photons deriving from the disk or the broad line region (EC)
is likely responsible for the emission at hard X- and $\gamma$-ray energies. 
Clarifying the  nature of the seed photons for this mechanism 
would give us insight into the physical cause(s) of
 the huge amplitude variations exhibited
by 3C~279 at the highest energies.

There have been several multi-wavelength observations of this blazar
 (Maraschi et al. 1994; Hartman et al. 1996; Wehrle et al. 1998). 
In particular, 3C~279 was monitored  with IUE 
and ROSAT for three weeks between December 1992 and January 1993, 
simultaneously with $\gamma$-ray observations by EGRET, and with coordinated
optical observations. At that epoch the intensity of 3C~279 was
 at a historical minimum at all measured wavelengths  above the sub-mm ones. 
 The variability amplitude with respect to the bright state of June 1991
increased with frequency from the IR to the UV. In fact it is possible that
the residual UV emission contains a substantial contribution from an 
accretion disk (see Pian et al. 1998b).
Regarding the second spectral component  the  variability amplitude was small 
in the soft (ROSAT) X-ray range but extremely large in $\gamma$-rays, larger than
in any other waveband. This showed that the inverse Compton emission had varied much
more than the synchrotron one, an effect expected in the SSC model. A similar
behaviour could be reproduced in the EC model only if the ambient photons
had varied in a correlated fashion with the electrons in the jet.

A second intensive multifrequency monitoring, based on  a 3-week coordinated
 program involving  CGRO, XTE and IUE observations,
 took place in Jan-Feb 1996
 (Wehrle et al. 1998). A huge flare was observed in $\gamma$-rays and a similar
but less extreme event was seen by XTE. The X and $\gamma$-ray peaks were
simultaneous 
within the one day uncertainty.
The IUE light curve at 2600 \AA\ (Fig.~1c) is reasonably well sampled during 
the first part of the campaign but not toward the end, when the flare
occurred and observations of 3C~279 were extended as Target of Opportunity.
 It shows a broad  minimum at $\sim$ Jan 25-26 followed by a rise
of almost a factor 2, but with a three day gap before and up to
 the $\gamma$-ray peak. Due to the faintness of the source and to the scattered
light
problem during the final stages of the IUE mission
systematic errors may affect some of the IUE observations.
Despite the non optimal sampling the IUE observations provide important constraints,
as discussed below.

\begin{figure}[h]
  \begin{center}
    \leavevmode
  \centerline{\epsfig{file=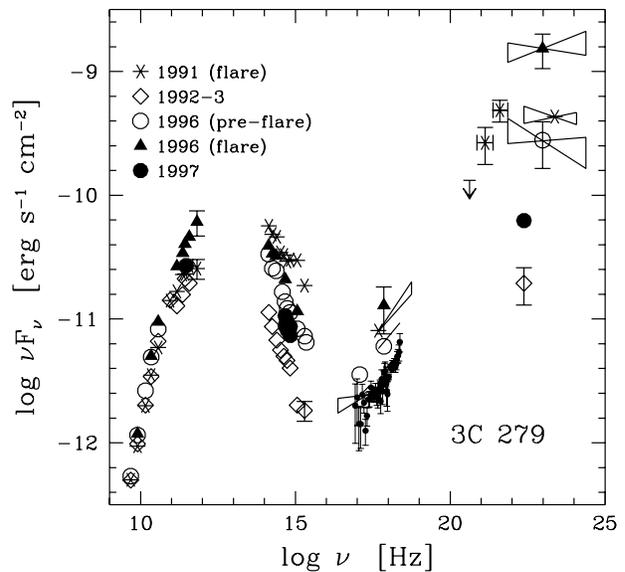,height=9cm}}
  \vspace{-1.cm}
  \end{center}
  \caption{\em Radio-to-$\gamma$-ray energy distribution of 3C~279 in pre-flare,
and flaring state in 1996 January-February (Wehrle et al. 1998).
The slope of the ASCA spectrum ($\alpha_\nu = 0.7$) has been reported
normalized to the RXTE point closest in time.  The EGRET best fit power-law spectra
referring to the January 16-30 (preflare) and February 4-6 (flare) periods are
shown, normalized at 0.4 GeV. Errors have been reported
only when they are bigger than the symbol size.
The UV, optical and near-IR data have been corrected for Galactic 
extinction.
For comparison, the SEDs in  
1991 June, in 1992 December -  93 January  are also shown
(see Maraschi et al. 1994). Preliminary results from observations performed in 1997
including the X-ray spectrum from the {\it Beppo}SAX satellite are shown as filled
small circles.}

%\label{fig:sample1}
\end{figure}

In Figure 3 the broad band energy distributions obtained at the flare peak
and in a preflare state (1996, Jan 25-26) are compared with those of the
discovery
epoch (June 1991) and with the low state of Jan 1993.
I also added new data obtained in 1997 including the X-ray spectrum
measured by {\it Beppo}SAX, referring  again to a rather low state.

Comparing the flare and preflare states it is clear that the high energy
SED (X-ray to $\gamma$-ray) is harder
at the flare peak, as implied by the larger amplitude of the $\gamma$-ray
variation.  From IR to  UV frequencies  the flare vs preflare variations are
much
smaller than in  X-rays and $\gamma$-rays. This has important consequences for
theoretical
models, as the relative variability in the synchrotron and
inverse Compton components can indicate the origin of the seed photons that may be
upscattered to the $\gamma$-ray band.
Between June 1991 
and December 1992--January 1993,
the $\gamma$-ray flux of 3C~279 varied approximately quadratically with respect
to the optical flux change. This is expected in an SSC scenario
(Maraschi, Ghisellini, \& Celotti 1992), but is also consistent
 with the external Compton (EC) model 
(Sikora 1994), provided that the bulk Lorentz factor of the jet varied.

The amplitude of the $\gamma$-ray variation during the 1996 outburst 
(factor 10) is more than the square of the observed IR--optical flux 
change (factor $\sim 1.5$) which is a severe difficulty for both the SSC and EC 
scenarios. Two possible ways out are the following: i) different emission zones
 could contribute to the IR-optical flux,
 diluting the intrinsic variation associated with the $\gamma$-ray flaring
region;  
 ii) the synchrotron peak corresponding to the
$\gamma$-ray peak may fall at frequencies lower than IR, 
where adequate variations could have 
occurred.
Thus we cannot completely rule out the SSC scenario. Similar arguments apply for the
"standard" EC scenario.

An interesting alternative which could resolve the above problem is
 the "relativistic mirror" model of Ghisellini and Madau (1996), combining
advantages
of both the SSC and EC models.
Here the seed photons are provided by rapidly varying 
 emission from a few clouds close to the jet and photoionized by an approaching
active blob in it. First, the photoionizing continuum is
beamed, and second, the electrons in
the jet see emission from the nearest cloud(s) as beamed. 
Thus there can be a double-beaming effect, leading to a more-than-quadratic
increase in $\gamma$-rays associated with variations in synchrotron emission
from the active blob.
The asymmetric shape of the X--ray curve, in which the decay seems
 faster than the rise, can be accommodated by the mirror model since
the inverse Compton emission drops sharply  once the active part of the jet
passes the broad line cloud(s).
The mirror model can be tested, even in the absence of any
available $\gamma$-ray observations, by monitoring
the Ly$\alpha$ line of 3C~279. A limited number of clouds, over a
limited velocity range, should respond  to the most rapidly varying (days)
 jet emission. However the amplitude may be small, therefore accurate
measurements are required as only possible with HST. The present limits
on the variability of the  Ly$\alpha$ emission in 3C~279 (Koratkar et al.
1998; see also Pian et al. 1998b) are not stringent enough to
constrain  the model significantly.

IUE and $\gamma$-ray observations of 3C 279 were repeated practically every year
since the $\gamma$-ray detection, including the two major campaigns discussed 
above.  In Figure 4 we present a comparison of the long term UV and
$\gamma$-ray light-curves.  
For simultaneous (within one day) IUE SWP and LWP observations, the UV flux
at 2000 \AA\ has been derived from a power-law fit of the dereddened
spectrum in the 1200-3000 \AA\ range (see Pian et al. 1998b;1998c). 
%interpolated between the spectral fluxes
%at 1750 \AA\ measured on dereddened SWP spectra (Koratkar et al. 1998) and at 
%2600 \AA\ measured on dereddened LWP spectra (Wehrle et al. 1998; 
%Pian et al. 1998c), assuming a power-law spectrum in the range 1200-3000 \AA.
For isolated SWP and LWP observations (Koratkar et al. 1998; Pian et al.
1998c; Wehrle et al. 1998), the flux at 2000 \AA\ has been
extrapolated from the spectrum using a typical power-law spectral index 
$\alpha \simeq 1.4$ ($F_\nu \propto \nu^{-\alpha}$) for high, medium, and
low states of 3C~279,
and $\alpha \simeq 1$ for very low states (Pian et al. 1998b).
The HST data are taken from Koratkar et al. (1998, see also Netzer et al. 
1993; Wehrle et al. 1998).
The figure clearly indicates that the two light 
curves are correlated. In Figure 5 we plot the UV fluxes (computed as
detailed above) vs the $\gamma$-ray fluxes
measured  within one month. 
%In order to increase the number of points we have used 
%both LWP and SWP fluxes computing from both an extrapolated 2000 A flux.
This confirms the correlation and shows that, at least on long timescales, 
the variability amplitude 
in the UV is smaller than that in $\gamma$-rays, but reasonably  compatible with
being the square root of it.  

\begin{figure}[h]
  \begin{center}
    \leavevmode
  \centerline{\epsfig{file=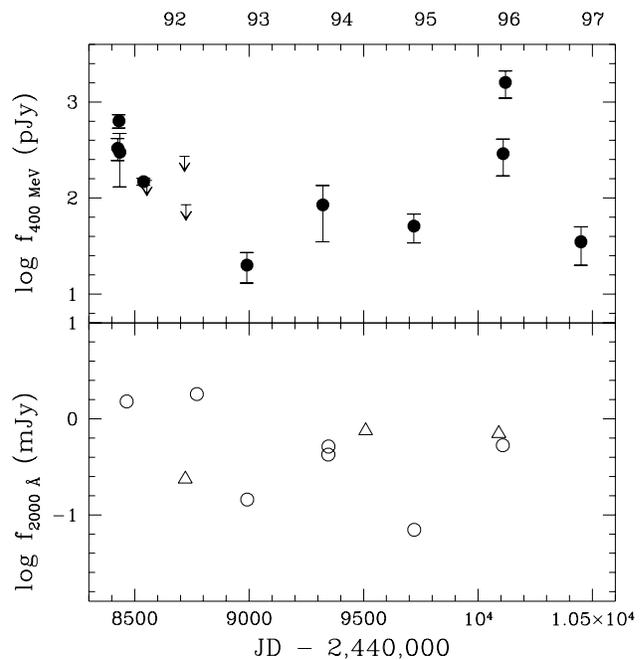,height=9cm}}
  \vspace{-0.4cm}
  \end{center}
  \caption{\em $\gamma$-ray (upper panel) and UV (lower panel) historical
light curves of 3C 279 from EGRET, IUE (open circles) and HST (open triangles)
archives. The uncertainties on the UV fluxes are smaller than the
symbol size and therefore not reported.}

%\label{fig:sample1}
\end{figure}

\begin{figure}[h]
  \begin{center}
    \leavevmode
  \centerline{\epsfig{file=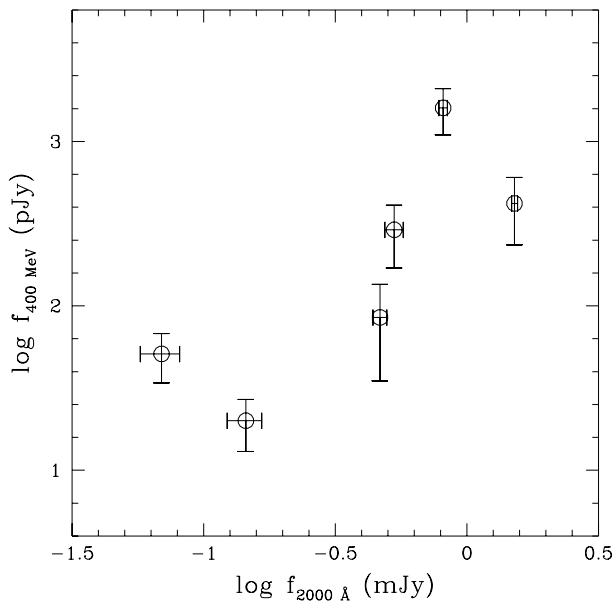,height=9cm}}
  \vspace{-0.5cm}
  \end{center}
  \caption{\em Correlation between $\gamma$-ray and UV flux in 3C 279. 
Only pairs of EGRET and IUE observations simultaneous within one month
have been considered.}

%\label{fig:sample1}
\end{figure}

\section{SUMMARY AND CONCLUSIONS}

Multifrequency studies of blazars seem to lead to a unified view
of the broad band continuum of the whole class. Two spectral components,
 each exhibiting a broad peak in the $\nu F_{\nu}$
representation are present in all objects. The two peak frequencies
are different in different objects but their ratio is approximately
constant. Irrespective of subclassifications, the blazar continua
may be described by a single spectral sequence whereby the values of the 
peak frequencies are fixed by the radio luminosity: higher luminosity 
objects peak at lower frequencies.

The overall regularity of the SEDs suggests that the same mechanisms 
(synchrotron and inverse Compton radiation) operate in all blazars,
albeit under gradually different physical conditions. The magnetic field,
the critical particle energy (corresponding to the maxima in the SED),
the importance of ambient vs. synchrotron photons in the inverse Compton
process may change gradually along the sequence (Ghisellini et al. 1997).

Multifrequency studies of objects at the "blue" end of the sequence,   
like PKS 2155-304, revealed the evolution of synchrotron
flares from X-rays to UV  frequencies,
indicating impulsive injection  of high energy particles
in the emission region. It is regrettable that IUE did not survive long 
enough to contribute to multifrequency campaigns including TeV observations, 
which only recently yielded positive detections of the brightest "blue"
blazars. 

For the "red" blazar 3C 279 multifrequency observations have shown
a clear long term correlation of the UV emission (due to the synchrotron
process) with the $\gamma$-ray emission (due to the inverse Compton
mechanism) thus providing strong support to the idea that the two spectral
components derive from the same population of relativistic electrons.
The short time-scale variability requires either a highly non linear
variation of the seed photons during a flare or dilution of the 
synchrotron flare by more stationary synchrotron emission from 
adjacent regions in the jet. 

\section*{ACKNOWLEDGEMENTS}

I thank Elena Pian for allowing the
presentation
of results in advance of publication and for help in the preparation
of the manuscript.

%\section{REFERENCES}

\end{document}